\documentclass[aps,prb,twocolumn,groupedaddress,amsmath,amssymb]{revtex4-1}

\usepackage{amsmath}             
\usepackage{amssymb}             
\usepackage{wasysym}             
\usepackage{color}               
\usepackage{setspace}            
\usepackage{graphicx}            
\usepackage{wrapfig}             
\usepackage[dvipsnames]{xcolor}  
\usepackage[user=PETER]{trkchg}    
\usepackage[userd=XSLI]{trkchgd}
\usepackage{array,booktabs}      
\usepackage{natbib}
\usepackage{units} 

\usepackage[font=footnotesize,labelfont=bf,labelsep=period,width=0.75\columnwidth]{caption}   
\usepackage[font=footnotesize,labelfont=bf,labelsep=period]{subcaption}                     
\DeclareCaptionSubType*[arabic]{figure}                                                     
\DeclareCaptionLabelFormat{subfiglabel}{Figure #2}                                          
\usepackage[capitalize]{cleveref}

\usepackage{xr}
\crefname{figure}{Fig.}{Figs.}
\Crefname{figure}{Figure}{Figures}
\crefname{table}{Tab.}{Tabs.}
\Crefname{table}{Table}{Tables}
\crefname{equation}{Eq.}{Eqs.}
\Crefname{equation}{Equation}{Equations}
\crefname{section}{Sec.}{Secs.}
\Crefname{section}{Section}{Sections}

\begin{document}

\title[CrI3 Skyrm]{Defect-Induced Magnetic Skyrmion in Two-Dimensional Chromium Tri-Iodide Monolayer}
\author{Ryan A. Beck}
\author{Lixin Lu}
\affiliation
{Department of Chemistry, University of Washington, Seattle, WA, 98195}
\author {Peter V. Sushko}
\affiliation
{Physical Sciences Division, Physical \& Computational Sciences Directorate, Pacific Northwest National Laboratory, Richland, WA, 99352}
\author{Xiaodong Xu}
\affiliation
{Department of Physics, University of Washington, Seattle, WA, 98195}
\author{Xiaosong Li}
\email{xsli@uw.edu}
\affiliation
{Department of Chemistry, University of Washington, Seattle, WA, 98195}

\date{\today}


%
%

\begin{abstract}
Chromium iodide monolayers, which have different magnetic properties in comparison to the bulk chromium iodide, have been shown to form skyrmionic states in applied electromagnetic fields or in Janus-layer devices.  In this work, we demonstrate that spin-canted solutions can be induced into monolayer chromium iodide by select substitution of iodide atoms with isovalent impurities.  Several concentrations and spatial configurations of halide substitutional defects are selected to probe the coupling between the local defect-induced geometric distortions and  orientation of chromium magnetic moments.  This work provides atomic-level insight into how atomically precise strain-engineering can be used to create and control complex magnetic patterns in chromium iodide layers and lays out the foundation for investigating the field- and geometric-dependent magnetic properties  in similar two-dimensional materials.
\end{abstract}

\maketitle

\section{Introduction}
Magnetic skyrmions are local whirls of the spins that both have a fixed chirality and do a full spin rotation.\cite{Cros17_17031} Isolated skyrmions can be treated as single particles and used in applications, such as system memory, radio frequency generators and filters, as well as spintronic devices.\cite{Cros17_17031,Lu18_055316}  
Typically, skyrmion formation in materials involves breaking their inversion symmetry which enables an asymmetric exchange interaction, the Dzyaloshinskii-Moriya interaction (DMI), through spin-orbit coupling.  The DMI interaction takes the form:\cite{Moria60_91}
\begin{equation}
\label{eq:DMI}
H_{DMI} = (S_1 \times S_2) \cdot \mathbf{d_{12}}
\end{equation}
where $S_1$ and $S_2$ are spins of two neighboring magnetic atoms and $\mathbf{d_{12}}$ is the corresponding Dzyaloshinskii-Moriya vector. If the exchange interaction between $S_1$ and $S_2$ is mediated by an anion, $\mathbf{d_{12}}$ can be written as $\mathbf{d_{12}} = \mathbf{r_1} \times \mathbf{r_2}$, where $\mathbf{r_1}$ and $\mathbf{r_2}$ link the anion with the two magnetic ions.\cite{Fernandez-Rossier17_035002} In CrX$_3$, this term is missing as the contributions associated with anionic pathways linking two neighboring Cr centers cancel out due to inversion symmetry.  
Commonly, skyrmion formation is induced by the application of an external electromagnetic field which, in the case of CrX$_3$, breaks the inversion symmetry of the system allowing the spins of the Cr centers to rotate. It has also been reported that Janus CrX$_3$ monolayers stabilize skyrmions.\cite{Lu18_055316,Lu18_054416,Cros17_17031,Bellaiche20_06404,Das19_232402}
Ideally, for practical use, materials would be able to realize skyrmions at high temperatures (those approaching room temperature) at zero, or very small, field strengths.\cite{Cros17_17031}

\begin{figure}\includegraphics[width=2.75in]{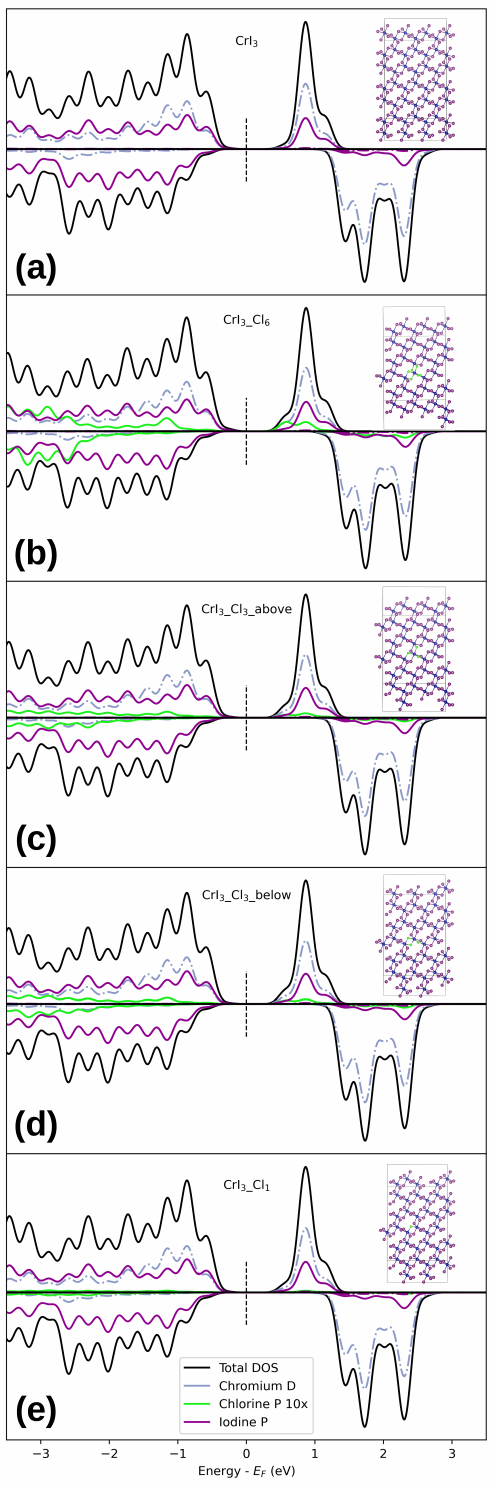}
\caption{The spin-resolved projected density of states for the monolayer CrI$_3$ systems (shown as insets to each plot).  The total DOS can be seen in each plot as the black line, the projected orbital contributions for the Cr-d (blue), I-p (purple), and Cl-p (green) atoms are plotted as well.  The pure CrI$_3$ monolayer system is shown at the top (a), with the halogen-replaced structures shown beneath: CrI$_3$\_Cl$_6$ (b), CrI$_3$\_Cl$_3$\_above (c), CrI$_3$\_Cl$_3$\_below (d), and CrI$_3$\_Cl$_1$ (e).
}
\label{fig:DOS}
\end{figure}

Chromium trihalides and, more broadly, MX$_3$ (where M is a transition metal) compounds are actively investigated for their unusual properties. They have a low synthesis and processing cost and
can be easily exfoliated to obtain few-layer materials. 
In chromium trihalide (CrX$_3$, X=Cl, Br, I) structures, the Cr$^{3+}$ ions are arranged into honeycomb lattices surrounded by six halogen anions giving rise to local octahedral symmetry.  The halogen atoms are each bound to two neighboring Cr centers.  At high temperature the layers stack with a monoclinic (space group C2/$m$) geometry.  At low temperature the layers stack with a rhombohedral (space group R$\overline{3}$) geometry.  The temperature at which this transformation occurs is dependent on the halogen (Cl=240 K, Br=420 K, and I = 210 K), and each bulk structure is known to have ferromagnetic moments between each Cr ion below their T$_c$ (Cl = 17 K, Br = 37 K, and I = 68 K).\cite{Mcguire15_612,Mcguire17_121}  Examples of the monolayer, hexagonal lattice and the bulk stacking can be seen in the insets to \cref{fig:DOS}.
  
CrI$_3$ bulk has been shown to exhibit the highest reported magnetic ordering temperature and anisotropy among the chromium trihalides.\cite{Mcguire15_612,Balke07_688,Wiesendanger14_696,Gregory50_5049,Xu17_270}
It has been shown that in the low-layer limit (number of layers $<\sim$10), MX$_3$ exhibit magnetic properties which appear to differ from their bulk properties.\cite{Xu17_270,Xu18_544,Xu19_1476,Xiao18_7658}
 Recently, mono- and bi-layers of chromium trihalide materials have been investigated for potential use in skyrmionic devices.\cite{Lu18_055316,Won20_166447,Cros17_17031,Bellaiche20_06404}  
 
 Monolayer CrI$_3$ is known to have long-range ferromagnetic character, in disagreement with the Mermin-Wagner theorem,\cite{Wagner66_1133,Lu18_055316} which is enabled through the magnetic anisotropy.\cite{Lu18_055316,Bellaiche18_57}  This magnetic anisotropy in CrI$_3$ arises from the spin-orbit coupling in iodine atoms, and favors a magnetic easy axis perpendicular to the atomic plane.\cite{Fernandez-Rossier17_035002,Mcguire17_121,Bellaiche18_57} 
Magnetic interactions between the chromium ions in a CrI$_3$ layer arise from a superexchange mechanism between the Cr 3$d$ orbitals and the I 5$p$ orbitals.\cite{Xiao18_7658,Fernandez-Rossier17_035002,Anderson50_350,Bellaiche18_57}  Geometric distortions in the CrI$_3$ layers can break the inversion symmetry, and thus induce finite  Dzyaloshinskii-Moriya interaction, ultimately leading to the appearance of skyrmionic ground states. In the case of an applied electric field oriented perpendicular  to the CrI$_3$ plane, the Cr$^{3+}$ and I$^-$ ions displace in the opposite directions in and out of the plane. These displacements change distances between the chromium and iodine atomic planes by as much as $\sim$3.6\% for the field magnitude of $\sim$0.2 V/nm, resulting in significant DMI effects.

In this work we investigate the formation of skyrmionic states via breaking the inversion  symmetry of the CrI$_3$ monolayer by substituting iodine atoms with chloride atoms (Cl$_{I}$). Unlike external electric fields, these defects produce localized atomic-scale distortions, which holds the promise of creating fine-tuned distortion patters and, accordingly,  may enable the formation of complex magnetic structures.

\section{Methodology}
Monolayer CrI$_3$ was represented using the isolated periodic slab model. The initial positions of atoms correspond to the bulk CrI$_3$ lattice as determined through single crystal X-ray diffraction at 90 K.\cite{Mcguire15_612} To find the optimal structure of the monolayer and its electronic properties, we performed ab initio simulations based on the density functional theory (DFT) as implemented in the Vienna \emph{ab initio} simulation program (VASP).\cite{Furthmuller96_15,Furthmuller96_11169} 
The projector augmented wave method and Perdew-Burk-Ernzerhof exchange correlation functional were used.\cite{Blochl94_17953,Ernzerhof96_3865} The calculations were performed in the spin-polarized mode. A plain-wave basis set was used with an energy cutoff of 600 eV, and the DFT-D3(0)\cite{Krieg10_132} method was used for dispersion correction. All internal coordinates were fully relaxed. It has been noted that the Hubbard U correction applied within the DFT+U method does not significantly impact the  results results of the calculations, and as such is not used herein.\cite{Fernandez-Rossier17_035002,Bellaiche18_57}

Two find the properties of pristine CrI$_3$, we used a supercell containing two chromium and six iodine atoms. A 3.5 nm vacuum gap in the off-plane direction (\emph{z}) was used to avoid the monolayer interactions with its periodic images. The Brillouin zone was sampled with a Monkhorst-Pack \emph{k}-point mesh of $6\times6\times2$.  The optimal lattice constant (\emph{a}$_0$) was found to be 6.929 \AA, which is in a close agreement with the experimentally observed value of 6.866 \AA.\cite{Mcguire15_612}  
  
The calculated Heisenberg isotropic symmetric exchange coefficient ($J = -2.99$ meV) and the magnetic anisotropy energy (MAE) (0.58 meV/Cr) are comparable with those previously reported for the CrI$_3$  systems (less than 10 layers): $J = -2.2$ meV and MAE = 0.65 meV/Cr.\cite{Fernandez-Rossier17_035002,Lu18_055316,Xu20_247201}

In order to examine the magnetic effects of Cl substitution, we used a  CrI$_3$ supercell of 1.8 nm $\times$ 3.1 nm $\times$ 2.0 nm. Several iodine atoms near a chromium center were replaced with Cl atoms, and the internal coordinates were optimized for each case.  
The cutoff for energy minimization with respect to the atomic coordinates was set to 10$^{-5}$ eV.  
With the optimized structure, a noncollinear (NC) spin calculation was run with the  self-consistent-field (SCF) convergence cutoff set to 10$^{-6}$ eV. The noncollinear wavefunction was used as a guess for the calculations incorporating spin-orbit coupling effects (NC-SOC) with SCF convergence set to 10$^{-9}$ eV in order to capture the energy cost of spin rotations which typically are on the order of 10$^{-6}\sim10^{-4}$ eV in magnitude.\cite{Lu18_055316}

\section{Results and Discussion}
\subsection{Geometric Distortion}\label{sec:geom}
In an attempt to introduce unique magnetic properties, chlorine atoms have been introduced into the CrI$_3$ monolayer in various configurations.  Chlorine was specifically chosen as it has been previously shown that bulk CrCl$_3$ exhibits an in-plane ferromagnetism,\cite{Mcguire15_612} which was thought could be taken advantage of to promote skyrmion formation without drastically altering the properties of the CrI$_3$ layer.  Also, of the CrX$_3$ systems commonly examined, Cl has the smallest atomic radius, and as such is thought to give rise to the largest geometric distortions.  Four CrI$_3$\_Cl$_x$ structures were examined, and can be viewed in the insets for \cref{fig:DOS}.   The four geometries considered have six iodine centers replaced around a chromium center (CrI\_Cl$_6$), three Cl replacing the iodine above the chromium atomic plane (CrI\_Cl$_3$\_above), three Cl replacing the iodine both above and below the chromium atomic plane (CrI\_Cl$_3$\_below), and one Cl replacing a single iodine atom (CrI\_Cl$_1$).

\begin{figure}
\includegraphics[width=3.25in]{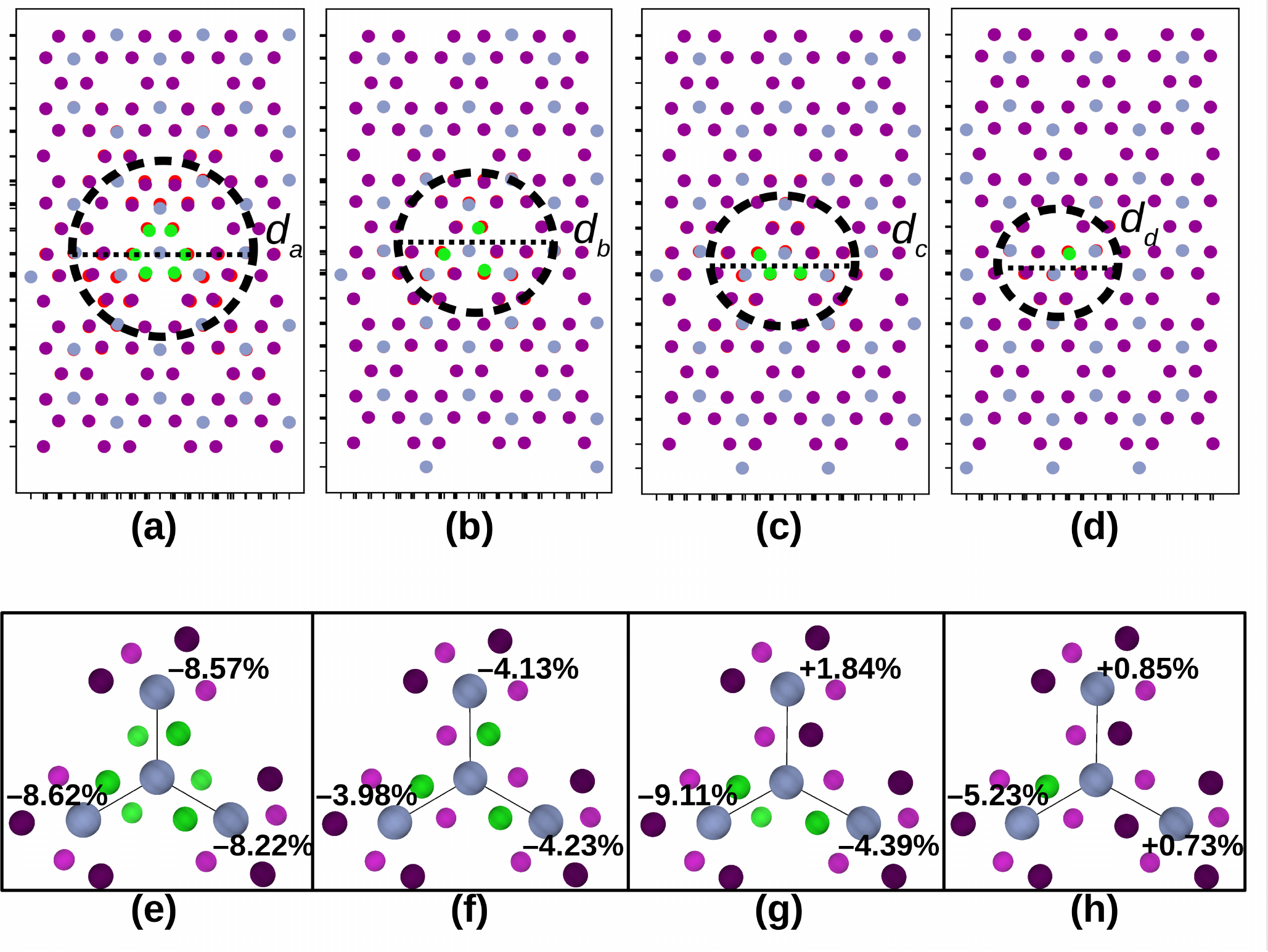}
\caption{The optimized geometries of the Cl substituted CrI$_3$ systems.  The initial position of the relevant atoms can be visualized by the red circles.  The finial geometries are shown with the gray circles representing Cr atoms, the purple circles representing iodine, and the green circles representing chlorine after optimization.  The geometries consist of CrI\_Cl$_6$ (\textbf{a}), CrI\_Cl$_3$\_above (\textbf{b}),  CrI\_Cl$_3$\_below (\textbf{c}), and CrI\_Cl$_1$ (\textbf{d}). 
\textbf{e-h} show the four central Cr atoms and bound halogen atoms.  The text shows the difference in the bond length between the central Cr atom and those neighboring from the pure CrI$_3$ system and the optimized, doped systems (shown as a $\pm$\% of the original Cr-Cr bond length).
}
\label{fig:CrIGeoms}
\end{figure}


As can be seen in \cref{fig:CrIGeoms}, replacing I atoms with Cl atoms results in a contraction of the Cr-Cl bonds (in comparison to the Cr-I bonds) as is expected given that Cl is both smaller and more electronegative than I.  In the case where six I atoms are replaced, the geometric distortions extend through the monolayer for $\sim$1.7 nm, similar to the extent of skyrmion formation which has been observed when CrI layers are exposed to a field.\cite{Lu18_055316} It should be noted that this size may be artificially constrained due to the size of the cluster used, as was also noted for the skyrmion formation by Liu \emph{et al.} (Ref. \citenum{Lu18_055316}).  
This distortion is shown in \cref{fig:CrIGeoms}a as the circle of diameter $d_a=1.7$ nm.  For systems where less Cl has been doped into the cell (in \cref{fig:CrIGeoms}b-d) the extent of the distortion caused is lessened.  CrI\_Cl$_3$\_above shows an extent of $d_b=1.2$ nm, and CrI\_Cl$_3$\_below and CrI\_Cl$_1$ ($d_c$ and $d_d$, respectively) show an extent of 1.0 nm.  The effects of the larger spatial extent of the distortion mean that there are additional chromium centers that no longer have the full octahedral symmetry surrounding them, thus giving rise to DM terms potentially leading to noncollinear spin solutions with no applied field.  It is of particular interest that both the CrI\_Cl$_6$ and CrI\_Cl$_3$\_above  systems have a symmetric distortion around the central Cr and thus the finial geometries are similar, while the CrI\_Cl$_3$\_below and CrI\_Cl$_1$ systems are less symmetric about the central Cr.  In the case of the CrI\_Cl$_3$\_below and CrI\_Cl$_1$ systems the central Cr is pulled slightly towards the Cr centers that are mediated by Cl atoms  leading to a significant distortion to the angles and bond lengths of the surrounding systems.

\begin{figure}
	\includegraphics[width=3.25in]{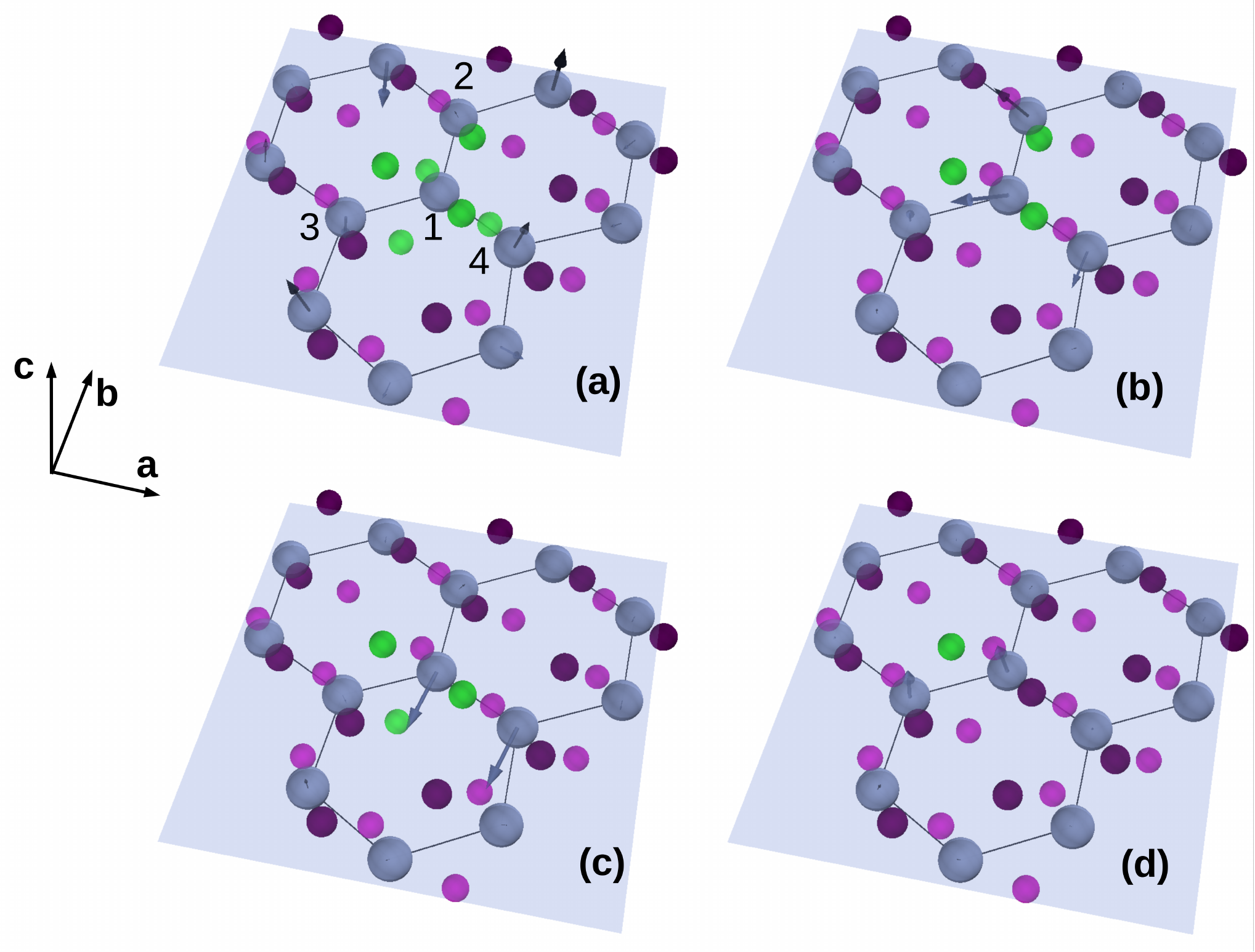}
	\caption{The DM vectors (in black, determined as by in Ref. \citenum{Keffer61_896}) on each Cr atom from the geometry optimized, Cl doped systems shown in \cref{fig:CrIGeoms}, CrI\_Cl$_6$ (a), CrI\_Cl$_3$\_above (b), CrI\_Cl$_3$\_below (c), and CrI\_Cl$_1$ (d).  Since the DM vectors are only non-zero where the inversion symmetry is broken, and the symmetry breaking is a localized effect, only the atoms that lie roughly within the black circle in \cref{fig:CrIGeoms} have been plotted.  The labeled atoms in (a) are the Cr atoms used in the four-state method to calculate the DM vectors that are reported in \cref{tab:DMvects}. }
	\label{fig:CrIDM}
\end{figure}

As a result of the geometric distortions, the inversion symmetry for the doped systems (except the CrI$_3$\_Cl$_6$ system) is broken.  Thus the DM terms become non-zero.  The DM vectors for the Cr atoms within the distorted areas marked in \cref{fig:CrIGeoms} are plotted in \cref{fig:CrIDM}.  This effect is localized to the areas with geometric distortion.  This can be seen in \cref{fig:CrIDM}(d), where the spatial extent of the geometry distortion is significantly less than in the other systems, by the fact that the Cr atoms on the far right (where there was little to no distortion) do not have applicable DM vectors.  As previously mentioned, the CrI$_3$\_Cl$_6$ system is able to maintain inversion symmetry around the central atom, and as such the DM contributions along each halide pathway cancel out, which is not the case for the other systems.

\begin{table*}
	\caption{DM vectors calculated using the four-state method of Refs. \citenum{Bellaiche18_57,Bellaiche20_06404}.  For the CrI$_3$\_Cl$_6$ and CrI$_3$\_Cl$_3$\_above, the three-fold symmetry surrounding the Cr center is maintained so only one vector is shown.  For the other systems, this symmetry is broken and the DM vectors between the central Cr and the three surrounding Cr centers are shown.  Cr centers are identified in \cref{fig:CrIDM}.}
	\begin{tabular}{l r r r r r}
		Pair & $\mathbf{D}_{ij\_X}$ (meV) & $\mathbf{D}_{ij\_Y}$ (meV) & $\mathbf{D}_{ij\_Z}$ (meV) & $|\mathbf{D}_{ij}|$ (meV) & $\nicefrac{|\mathbf{D}_{ij}|}{J_{ij}}$ (meV)\\\hline\hline
		CrI$_3$\_Cl$_6$ Cr$_1$ Cr$_2$ & $-$0.28 & 0.34 & $-$0.22	&	0.49 & 7.24\\\hline
		CrI$_3$\_Cl$_3$\_above Cr$_1$ Cr$_2$ & $-$1.23 & $-$0.79 & $-$1.48	&	2.08 & 2.42\\\hline
		CrI$_3$\_Cl$_3$\_below Cr$_1$ Cr$_2$	& 0.22	& 0.72	&	1.02	& 1.27	& 0.92\\
		CrI$_3$\_Cl$_3$\_below Cr$_1$ Cr$_3$	& 0.22 & 0.81 & 0.21 	&	0.87 & 0.49\\
		CrI$_3$\_Cl$_3$\_below Cr$_1$ Cr$_4$	& 0.26 & $-$0.02 & $-$0.77	&	0.81 & 13.55\\\hline
		CrI$_3$\_Cl$_1$ Cr$_1$ Cr$_2$	& $-$0.38 & $-$0.16 & 1.15	&	1.22 &2.91\\
		CrI$_3$\_Cl$_1$ Cr$_1$ Cr$_3$	& 0.22	&	0.40	&	$-$0.56	&	0.72 &2.77\\
		CrI$_3$\_Cl$_1$ Cr$_1$ Cr$_4$	&	$-$0.19	&	$-$0.53	&	0.08	&	0.13 &0.49\\\hline\hline
	\end{tabular}
	\label{tab:DMvects}
\end{table*}

\subsection{Formation of Spin Bubble}\label{sec:spin}
\begin{figure}
	\includegraphics[width=3.25in]{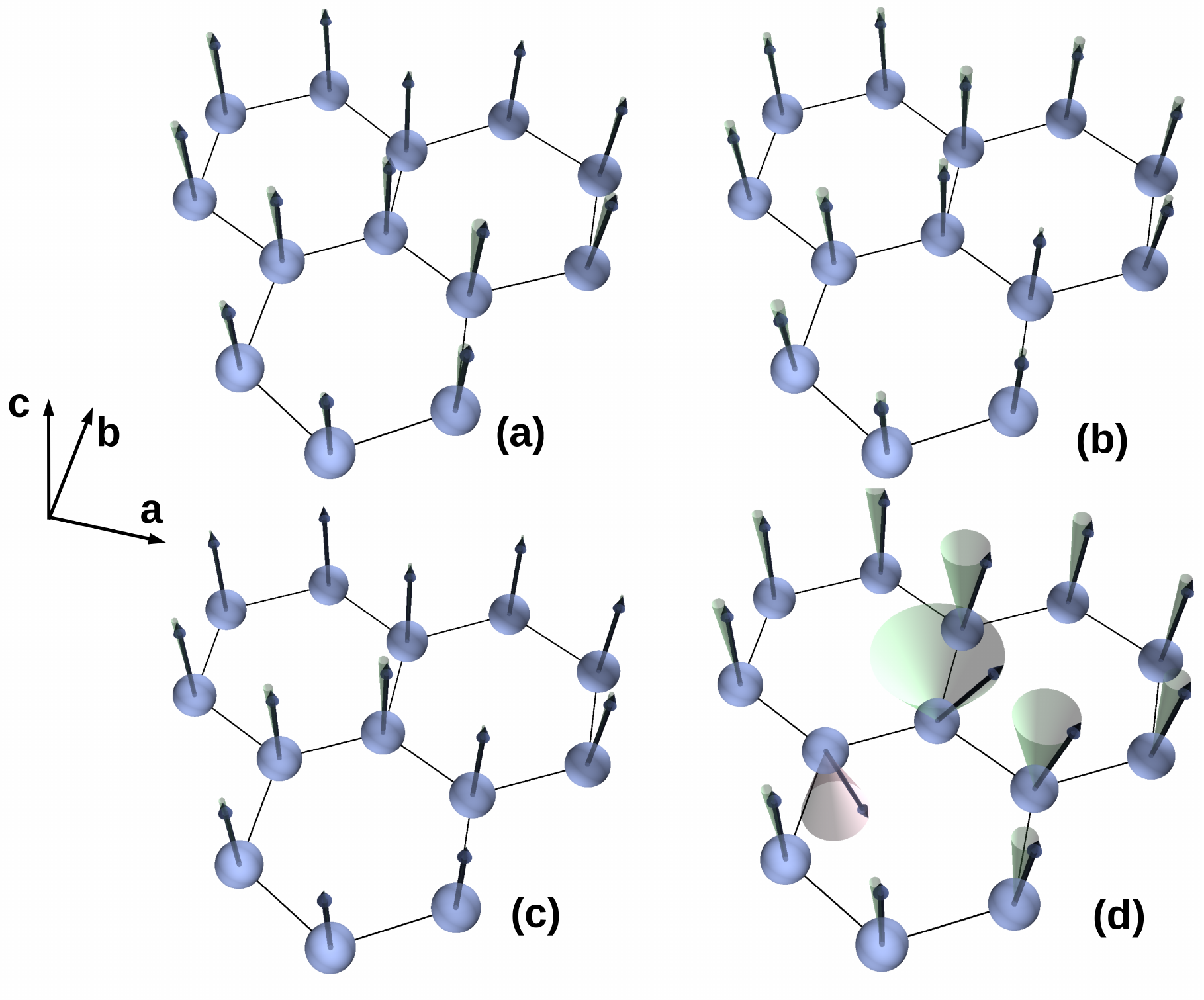}
	\caption{The canted spin states for each of the systems resulting from the presence of the doped Cl shown by black vectors.  The difference between the canted spins and the collinear spins perpendicular to the crystal plane are shown by the cones (where green cones are spins above the layer and red cones are spins beneath the layer).}
	\label{fig:CrISkrm}
\end{figure}

Magnetic skyrmions are local whirls of the spins that both have a fixed chirality and do a full spin rotation.\cite{Cros17_17031}  Unfortunately none of the geometries attempted in this study are able to realize a full symmetric skyrmion at zero field. Instead, spin bubbles can be observed in the CrI$_3$ monolayers arising from spin noncollinear solutions.  In the case of the CrI$_3$\_Cl$_1$ system a spin bubble solution was stabilized as can be seen in \cref{fig:CrISkrm}.  Plotted in \cref{fig:CrISkrm} are the atomic clusters shown in \cref{fig:CrIDM} with the magnetization vectors for each Cr atom plotted.  It can be noted that in the case of the CrI$_3$\_Cl$_3$ systems, the spin bubble is unable to manifest and the ferromagnetic solution persists.  

In the case of the CrI$_3$\_Cl$_1$ system, the stabilization by the formation of the spin bubble over the ferromagnetic solution is 25.7 meV/Cr.  This energetic difference is larger than what has been previously observed  as the stabilization energy from the formation of skyrmions through the use of external fields ($\sim$3 meV/Cr) is much larger than the cost of a spin rotation ($<$1 meV/Cr).\cite{Lu18_055316}  Thus similar to the skyrmions formed through the application of external fields, the spin bubbles are topologically-protected spin configurations.
It is interesting to note that while able to form in the CrI$_3$\_Cl$_1$  system, a spin bubble configuration is not manifested in the CrI$_3$\_Cl$_3$ \_below system.  Even though, in the CrI$_3$\_Cl$_3$ \_below system, there is both a lack of symmetry in the geometry (as in either the CrI$_3$\_Cl$_6$ or CrI$_3$\_Cl$_3$\_above cases), and there exists a pathway with only a single Cl atom that has a similar geometric distortion to the CrI$_3$\_Cl$_1$  system.  This can be seen in \cref{fig:CrIGeoms} g and h, with a 4.39\% reduction in Cr-Cr distance for the CrI$_3$\_Cl$_3$\_below compared to a 5.23\% reduction for the CrI$_3$\_Cl$_1$.  It was previously noted that a minimum field was required in order to induce skyrmions which introduced a distortion between the Cr and I layers in the monolayer CrI$_3$.\cite{Lu18_055316}  
Measuring geometric changes for the systems finds that the CrI$_3$\_Cl$_3$\_below system has a distortion of 2.2\% while the CrI$_3$\_Cl$_1$ system has a distortion of 3.1\%.  Given that the distortion of the CrI$_3$\_Cl$_3$\_below system is less than the previously determined minimum may contribute to why a magnetic bubble state is unable to manifest even though there are DM vectors present.

In order to gain insights into the formation of the spin bubble in the CrI$_3$\_Cl$_1$ system, the DMI vectors were calculated for the Cr atoms shown in \cref{fig:CrIDM}.  
The numerical values for the DM vectors have been tabulated in \cref{tab:DMvects} and calculated through the four-state method detailed in Refs. \citenum{Bellaiche18_57,Bellaiche20_06404}.
For the CrI$_3$\_Cl$_6$ and CrI$_3$\_Cl$_3$\_above systems only the Cr$_1$ and Cr$_2$ pair were calculated due to the presence of the rotational symmetry. For the other systems, the DMI vectors for the nearest neighbors to the central Cr were all explicitly calculated.  Previous work has identified that skyrmions are more likely to form when values of $\nicefrac{D_{ij}}{J_{ij}}$ (where $D_{12} = |\mathbf{D}_{ij}|$) are between 0.1-0.2 meV.\cite{Bellaiche20_06404,Sampaio13_152}  Larger values of the $\nicefrac{D_{ij}}{J_{ij}}$ value favors faster spin rotations about $\mathbf{D}_{ij}$ leading to smaller skyrmions.\cite{Sampaio13_152}  Our values for $\nicefrac{D_{ij}}{J_{ij}}$ are significantly larger than those reported for previous CrI$_3$ systems.  This is most likely due to the increased localized distortion surrounding the Cr centers.  The values for $|\mathbf{D}_{ij}|$ are also larger than those previously reported,\cite{Bellaiche20_06404} though it should be noted that large values for the DMI vectors have been shown in previous studies and will often introduce other spin configurations than skyrmions.\cite{Sampaio13_152}  
Although the CrI$_3$\_Cl$_1$ (Cr$_1$,Cr$_3$) and CrI$_3$\_Cl$_3$\_below (Cr$_1$,Cr$_4$) systems are close in geometry and as such expected to both exhibit spin canted solutions, the CrI$_3$\_Cl$_3$\_below system has a $\nicefrac{D_{1,4}}{J}$ value (tabulated in \cref{tab:DMvects}) that is too large to manifest a stable spin whirl solution at zero field.  In contrast, while the value for $\nicefrac{D_{ij}}{J}$ is larger than in other reports of skyrmion-forming monolayer chromium iodide systems, a spin canted solution is still able to manifest for the CrI$_3$\_Cl$_1$ system.  Since selective doping was able to introduce a spin-canted solution additional studies on alternate geometries and the field effects can be carried out.


\section{Conclusion}
We show that upon substituting selected Iodine atoms in a CrI$_3$ monolayer with Cl atoms, localized spin-bubble states form.  These spin states arise from the lattice distortions induced by the ionic radii mismatch between the host and the defect species. While distortions are driven by the difference in the X-Cr bond lengths, the interactions between these distortions can induced long-range directional lattice polarization that may enable coupling between spatially-separated spin-bubble states.  It was noted that, when these spin-bubble systems formed, they were topologically protected as they were significantly more stable than either the cost of a spin-flip or the stabilization noted by electric-field induced skyrmions.  This work provided an important step toward manifesting skyrmions at conditions that would be useful for spintronic applications by potentially reducing the field required and increasing the operating temperature through controlled doping of CrI$_3$ monolayers.  

\section{acknowledgement}
The work was supported by the Northwest Institute for Materials Physics, Chemistry, and Technology (NW IMPACT), and the University of Washington Molecular Engineering Materials Center (DMR-1719797).  This work was facilitated though the use of advanced computational, storage, and networking infrastructure provided by the Hyak supercomputer system and funded by the STF at the University of Washington.

%

\newpage
\clearpage




\newpage 
\clearpage
\bibliography{Journal_Short_Name,Li_Group_References,BECK_REFERENCES,Gamelin_Group_References} 

\begin{thebibliography}{27}%
\makeatletter
\providecommand \@ifxundefined [1]{%
 \@ifx{#1\undefined}
}%
\providecommand \@ifnum [1]{%
 \ifnum #1\expandafter \@firstoftwo
 \else \expandafter \@secondoftwo
 \fi
}%
\providecommand \@ifx [1]{%
 \ifx #1\expandafter \@firstoftwo
 \else \expandafter \@secondoftwo
 \fi
}%
\providecommand \natexlab [1]{#1}%
\providecommand \enquote  [1]{``#1''}%
\providecommand \bibnamefont  [1]{#1}%
\providecommand \bibfnamefont [1]{#1}%
\providecommand \citenamefont [1]{#1}%
\providecommand \href@noop [0]{\@secondoftwo}%
\providecommand \href [0]{\begingroup \@sanitize@url \@href}%
\providecommand \@href[1]{\@@startlink{#1}\@@href}%
\providecommand \@@href[1]{\endgroup#1\@@endlink}%
\providecommand \@sanitize@url [0]{\catcode `\\12\catcode `\$12\catcode
  `\&12\catcode `\#12\catcode `\^12\catcode `\_12\catcode `\%12\relax}%
\providecommand \@@startlink[1]{}%
\providecommand \@@endlink[0]{}%
\providecommand \url  [0]{\begingroup\@sanitize@url \@url }%
\providecommand \@url [1]{\endgroup\@href {#1}{\urlprefix }}%
\providecommand \urlprefix  [0]{URL }%
\providecommand \Eprint [0]{\href }%
\providecommand \doibase [0]{http://dx.doi.org/}%
\providecommand \selectlanguage [0]{\@gobble}%
\providecommand \bibinfo  [0]{\@secondoftwo}%
\providecommand \bibfield  [0]{\@secondoftwo}%
\providecommand \translation [1]{[#1]}%
\providecommand \BibitemOpen [0]{}%
\providecommand \bibitemStop [0]{}%
\providecommand \bibitemNoStop [0]{.\EOS\space}%
\providecommand \EOS [0]{\spacefactor3000\relax}%
\providecommand \BibitemShut  [1]{\csname bibitem#1\endcsname}%
\let\auto@bib@innerbib\@empty
\bibitem [{\citenamefont {Fert}\ \emph {et~al.}(2017)\citenamefont {Fert},
  \citenamefont {Reyren},\ and\ \citenamefont {Cros}}]{Cros17_17031}%
  \BibitemOpen
  \bibfield  {author} {\bibinfo {author} {\bibfnamefont {A.}~\bibnamefont
  {Fert}}, \bibinfo {author} {\bibfnamefont {N.}~\bibnamefont {Reyren}}, \ and\
  \bibinfo {author} {\bibfnamefont {V.}~\bibnamefont {Cros}},\ }\href {\doibase
  10.1038/natrevmats.2017.31} {\bibfield  {journal} {\bibinfo  {journal} {Nat.
  Rev.}\ }\textbf {\bibinfo {volume} {2}},\ \bibinfo {pages} {17031} (\bibinfo
  {year} {2017})}\BibitemShut {NoStop}%
\bibitem [{\citenamefont {Liu}\ \emph {et~al.}(2018{\natexlab{a}})\citenamefont
  {Liu}, \citenamefont {Shi}, \citenamefont {Mo},\ and\ \citenamefont
  {Lu}}]{Lu18_055316}%
  \BibitemOpen
  \bibfield  {author} {\bibinfo {author} {\bibfnamefont {J.}~\bibnamefont
  {Liu}}, \bibinfo {author} {\bibfnamefont {M.}~\bibnamefont {Shi}}, \bibinfo
  {author} {\bibfnamefont {P.}~\bibnamefont {Mo}}, \ and\ \bibinfo {author}
  {\bibfnamefont {J.}~\bibnamefont {Lu}},\ }\href {\doibase 10.1063/1.5030441}
  {\bibfield  {journal} {\bibinfo  {journal} {AIP Advances}\ }\textbf {\bibinfo
  {volume} {8}},\ \bibinfo {pages} {055316} (\bibinfo {year}
  {2018}{\natexlab{a}})}\BibitemShut {NoStop}%
\bibitem [{\citenamefont {Moriya}(1960)}]{Moria60_91}%
  \BibitemOpen
  \bibfield  {author} {\bibinfo {author} {\bibfnamefont {T.}~\bibnamefont
  {Moriya}},\ }\href@noop {} {\bibfield  {journal} {\bibinfo  {journal} {Phys.
  Rev.}\ }\textbf {\bibinfo {volume} {120}},\ \bibinfo {pages} {91} (\bibinfo
  {year} {1960})}\BibitemShut {NoStop}%
\bibitem [{\citenamefont {Lado}\ and\ \citenamefont
  {Fern\'andez-Rossier}(2017)}]{Fernandez-Rossier17_035002}%
  \BibitemOpen
  \bibfield  {author} {\bibinfo {author} {\bibfnamefont {J.~L.}\ \bibnamefont
  {Lado}}\ and\ \bibinfo {author} {\bibfnamefont {J.}~\bibnamefont
  {Fern\'andez-Rossier}},\ }\href@noop {} {\bibfield  {journal} {\bibinfo
  {journal} {2D Mater.}\ }\textbf {\bibinfo {volume} {4}},\ \bibinfo {pages}
  {035002} (\bibinfo {year} {2017})}\BibitemShut {NoStop}%
\bibitem [{\citenamefont {Liu}\ \emph {et~al.}(2018{\natexlab{b}})\citenamefont
  {Liu}, \citenamefont {Shi}, \citenamefont {Lu},\ and\ \citenamefont
  {Anantram}}]{Lu18_054416}%
  \BibitemOpen
  \bibfield  {author} {\bibinfo {author} {\bibfnamefont {J.}~\bibnamefont
  {Liu}}, \bibinfo {author} {\bibfnamefont {M.}~\bibnamefont {Shi}}, \bibinfo
  {author} {\bibfnamefont {J.}~\bibnamefont {Lu}}, \ and\ \bibinfo {author}
  {\bibfnamefont {M.~P.}\ \bibnamefont {Anantram}},\ }\href {\doibase
  10.1103/PhysRevB.97.054416} {\bibfield  {journal} {\bibinfo  {journal} {Phys.
  Rev. B}\ }\textbf {\bibinfo {volume} {97}},\ \bibinfo {pages} {054416}
  (\bibinfo {year} {2018}{\natexlab{b}})}\BibitemShut {NoStop}%
\bibitem [{\citenamefont {Xu}\ \emph {et~al.}(2020)\citenamefont {Xu},
  \citenamefont {Feng}, \citenamefont {Prokhorenko}, \citenamefont {Nahas},
  \citenamefont {Xiang},\ and\ \citenamefont {Bellaiche}}]{Bellaiche20_06404}%
  \BibitemOpen
  \bibfield  {author} {\bibinfo {author} {\bibfnamefont {C.}~\bibnamefont
  {Xu}}, \bibinfo {author} {\bibfnamefont {J.}~\bibnamefont {Feng}}, \bibinfo
  {author} {\bibfnamefont {S.}~\bibnamefont {Prokhorenko}}, \bibinfo {author}
  {\bibfnamefont {Y.}~\bibnamefont {Nahas}}, \bibinfo {author} {\bibfnamefont
  {H.}~\bibnamefont {Xiang}}, \ and\ \bibinfo {author} {\bibfnamefont
  {L.}~\bibnamefont {Bellaiche}},\ }\href {\doibase
  10.1103/PhysRevB.101.060404} {\bibfield  {journal} {\bibinfo  {journal}
  {Phys. Rev. B}\ ,\ \bibinfo {pages} {060404}} (\bibinfo {year}
  {2020})}\BibitemShut {NoStop}%
\bibitem [{\citenamefont {Behera}\ \emph {et~al.}(2019)\citenamefont {Behera},
  \citenamefont {Chowdhury},\ and\ \citenamefont {Das}}]{Das19_232402}%
  \BibitemOpen
  \bibfield  {author} {\bibinfo {author} {\bibfnamefont {A.~K.}\ \bibnamefont
  {Behera}}, \bibinfo {author} {\bibfnamefont {S.}~\bibnamefont {Chowdhury}}, \
  and\ \bibinfo {author} {\bibfnamefont {S.~R.}\ \bibnamefont {Das}},\ }\href
  {\doibase 10.1063/1.5096782} {\bibfield  {journal} {\bibinfo  {journal}
  {Appl. Phys. Lett.}\ ,\ \bibinfo {pages} {232402}} (\bibinfo {year}
  {2019})}\BibitemShut {NoStop}%
\bibitem [{\citenamefont {Huang}\ \emph {et~al.}(2017)\citenamefont {Huang},
  \citenamefont {Clark}, \citenamefont {Navarro-Moratalla}, \citenamefont
  {Klien}, \citenamefont {Cheng}, \citenamefont {Seyler}, \citenamefont
  {Zhong}, \citenamefont {Schmidgall}, \citenamefont {McGuire}, \citenamefont
  {Cobden}, \citenamefont {Yao}, \citenamefont {Xiao}, \citenamefont
  {Jarillo-Herrero},\ and\ \citenamefont {Xu}}]{Xu17_270}%
  \BibitemOpen
  \bibfield  {author} {\bibinfo {author} {\bibfnamefont {B.}~\bibnamefont
  {Huang}}, \bibinfo {author} {\bibfnamefont {G.}~\bibnamefont {Clark}},
  \bibinfo {author} {\bibfnamefont {E.}~\bibnamefont {Navarro-Moratalla}},
  \bibinfo {author} {\bibfnamefont {D.~R.}\ \bibnamefont {Klien}}, \bibinfo
  {author} {\bibfnamefont {R.}~\bibnamefont {Cheng}}, \bibinfo {author}
  {\bibfnamefont {K.~L.}\ \bibnamefont {Seyler}}, \bibinfo {author}
  {\bibfnamefont {D.}~\bibnamefont {Zhong}}, \bibinfo {author} {\bibfnamefont
  {E.}~\bibnamefont {Schmidgall}}, \bibinfo {author} {\bibfnamefont {M.~A.}\
  \bibnamefont {McGuire}}, \bibinfo {author} {\bibfnamefont {D.~H.}\
  \bibnamefont {Cobden}}, \bibinfo {author} {\bibfnamefont {W.}~\bibnamefont
  {Yao}}, \bibinfo {author} {\bibfnamefont {D.}~\bibnamefont {Xiao}}, \bibinfo
  {author} {\bibfnamefont {P.}~\bibnamefont {Jarillo-Herrero}}, \ and\ \bibinfo
  {author} {\bibfnamefont {X.}~\bibnamefont {Xu}},\ }\href {\doibase
  10.1038/nature22391} {\bibfield  {journal} {\bibinfo  {journal} {Nature}\
  }\textbf {\bibinfo {volume} {546}},\ \bibinfo {pages} {270} (\bibinfo {year}
  {2017})}\BibitemShut {NoStop}%
\bibitem [{\citenamefont {Huang}\ \emph {et~al.}(2018)\citenamefont {Huang},
  \citenamefont {genevieve Clark}, \citenamefont {Klein}, \citenamefont
  {MacNeil}, \citenamefont {Navarro-Moratalla}, \citenamefont {Seyler},
  \citenamefont {Wilson}, \citenamefont {McGuire}, \citenamefont {Cobden},
  \citenamefont {Xiao}, \citenamefont {yao}, \citenamefont {Jarillo-Herrero},\
  and\ \citenamefont {Xu}}]{Xu18_544}%
  \BibitemOpen
  \bibfield  {author} {\bibinfo {author} {\bibfnamefont {B.}~\bibnamefont
  {Huang}}, \bibinfo {author} {\bibnamefont {genevieve Clark}}, \bibinfo
  {author} {\bibfnamefont {D.~R.}\ \bibnamefont {Klein}}, \bibinfo {author}
  {\bibfnamefont {D.}~\bibnamefont {MacNeil}}, \bibinfo {author} {\bibfnamefont
  {E.}~\bibnamefont {Navarro-Moratalla}}, \bibinfo {author} {\bibfnamefont
  {K.~L.}\ \bibnamefont {Seyler}}, \bibinfo {author} {\bibfnamefont
  {N.}~\bibnamefont {Wilson}}, \bibinfo {author} {\bibfnamefont {M.~A.}\
  \bibnamefont {McGuire}}, \bibinfo {author} {\bibfnamefont {D.~H.}\
  \bibnamefont {Cobden}}, \bibinfo {author} {\bibfnamefont {D.}~\bibnamefont
  {Xiao}}, \bibinfo {author} {\bibfnamefont {W.}~\bibnamefont {yao}}, \bibinfo
  {author} {\bibfnamefont {P.}~\bibnamefont {Jarillo-Herrero}}, \ and\ \bibinfo
  {author} {\bibfnamefont {X.}~\bibnamefont {Xu}},\ }\href@noop {} {\bibfield
  {journal} {\bibinfo  {journal} {Nat. Nanotechnol.}\ }\textbf {\bibinfo
  {volume} {13}},\ \bibinfo {pages} {544} (\bibinfo {year} {2018})}\BibitemShut
  {NoStop}%
\bibitem [{\citenamefont {Song}\ \emph {et~al.}(2019)\citenamefont {Song},
  \citenamefont {Fei}, \citenamefont {Yankowitz}, \citenamefont {Lin},
  \citenamefont {Jiang}, \citenamefont {Hwangbo}, \citenamefont {Zhang},
  \citenamefont {Sun}, \citenamefont {Taniguchi}, \citenamefont {Watanabe},
  \citenamefont {McGuire}, \citenamefont {Graf}, \citenamefont {Cao},
  \citenamefont {Chu}, \citenamefont {Cobden}, \citenamefont {Dean},
  \citenamefont {Xiao},\ and\ \citenamefont {Xu}}]{Xu19_1476}%
  \BibitemOpen
  \bibfield  {author} {\bibinfo {author} {\bibfnamefont {T.}~\bibnamefont
  {Song}}, \bibinfo {author} {\bibfnamefont {Z.}~\bibnamefont {Fei}}, \bibinfo
  {author} {\bibfnamefont {M.}~\bibnamefont {Yankowitz}}, \bibinfo {author}
  {\bibfnamefont {Z.}~\bibnamefont {Lin}}, \bibinfo {author} {\bibfnamefont
  {Q.}~\bibnamefont {Jiang}}, \bibinfo {author} {\bibfnamefont
  {K.}~\bibnamefont {Hwangbo}}, \bibinfo {author} {\bibfnamefont
  {Q.}~\bibnamefont {Zhang}}, \bibinfo {author} {\bibfnamefont
  {B.}~\bibnamefont {Sun}}, \bibinfo {author} {\bibfnamefont {T.}~\bibnamefont
  {Taniguchi}}, \bibinfo {author} {\bibfnamefont {K.}~\bibnamefont {Watanabe}},
  \bibinfo {author} {\bibfnamefont {M.~A.}\ \bibnamefont {McGuire}}, \bibinfo
  {author} {\bibfnamefont {D.}~\bibnamefont {Graf}}, \bibinfo {author}
  {\bibfnamefont {T.}~\bibnamefont {Cao}}, \bibinfo {author} {\bibfnamefont
  {J.-H.}\ \bibnamefont {Chu}}, \bibinfo {author} {\bibfnamefont {D.~H.}\
  \bibnamefont {Cobden}}, \bibinfo {author} {\bibfnamefont {C.~R.}\
  \bibnamefont {Dean}}, \bibinfo {author} {\bibfnamefont {D.}~\bibnamefont
  {Xiao}}, \ and\ \bibinfo {author} {\bibfnamefont {X.}~\bibnamefont {Xu}},\
  }\href {\doibase 10.1038/s41563-019-0505-2} {\bibfield  {journal} {\bibinfo
  {journal} {Nat. Mater.}\ ,\ \bibinfo {pages} {1476}} (\bibinfo {year}
  {2019})}\BibitemShut {NoStop}%
\bibitem [{\citenamefont {Sivadas}\ \emph {et~al.}(2018)\citenamefont
  {Sivadas}, \citenamefont {Okamoto}, \citenamefont {Xu}, \citenamefont
  {Fennie},\ and\ \citenamefont {Xiao}}]{Xiao18_7658}%
  \BibitemOpen
  \bibfield  {author} {\bibinfo {author} {\bibfnamefont {N.}~\bibnamefont
  {Sivadas}}, \bibinfo {author} {\bibfnamefont {S.}~\bibnamefont {Okamoto}},
  \bibinfo {author} {\bibfnamefont {X.}~\bibnamefont {Xu}}, \bibinfo {author}
  {\bibfnamefont {C.~J.}\ \bibnamefont {Fennie}}, \ and\ \bibinfo {author}
  {\bibfnamefont {D.}~\bibnamefont {Xiao}},\ }\href {\doibase
  10.1021/acs.nanolett.8b03321} {\bibfield  {journal} {\bibinfo  {journal}
  {Nano Lett.}\ }\textbf {\bibinfo {volume} {18}},\ \bibinfo {pages} {7658}
  (\bibinfo {year} {2018})}\BibitemShut {NoStop}%
\bibitem [{\citenamefont {McGuire}\ \emph {et~al.}(2015)\citenamefont
  {McGuire}, \citenamefont {hemant Dixit}, \citenamefont {Cooper},\ and\
  \citenamefont {Sales}}]{Mcguire15_612}%
  \BibitemOpen
  \bibfield  {author} {\bibinfo {author} {\bibfnamefont {M.~A.}\ \bibnamefont
  {McGuire}}, \bibinfo {author} {\bibnamefont {hemant Dixit}}, \bibinfo
  {author} {\bibfnamefont {V.~R.}\ \bibnamefont {Cooper}}, \ and\ \bibinfo
  {author} {\bibfnamefont {B.~C.}\ \bibnamefont {Sales}},\ }\href {\doibase
  10.1021/cm5042421} {\bibfield  {journal} {\bibinfo  {journal} {Comp. Mater.}\
  }\textbf {\bibinfo {volume} {27}},\ \bibinfo {pages} {612} (\bibinfo {year}
  {2015})}\BibitemShut {NoStop}%
\bibitem [{\citenamefont {Felser}\ \emph {et~al.}(2007)\citenamefont {Felser},
  \citenamefont {Fecher},\ and\ \citenamefont {Balke}}]{Balke07_688}%
  \BibitemOpen
  \bibfield  {author} {\bibinfo {author} {\bibfnamefont {C.}~\bibnamefont
  {Felser}}, \bibinfo {author} {\bibfnamefont {G.~H.}\ \bibnamefont {Fecher}},
  \ and\ \bibinfo {author} {\bibfnamefont {B.}~\bibnamefont {Balke}},\ }\href
  {\doibase 10.1002/anie.200601815} {\bibfield  {journal} {\bibinfo  {journal}
  {Angew. Chem.}\ }\textbf {\bibinfo {volume} {46}},\ \bibinfo {pages} {688}
  (\bibinfo {year} {2007})}\BibitemShut {NoStop}%
\bibitem [{\citenamefont {Behin-Aein}\ \emph {et~al.}(2014)\citenamefont
  {Behin-Aein}, \citenamefont {Wang},\ and\ \citenamefont
  {Wiesendanger}}]{Wiesendanger14_696}%
  \BibitemOpen
  \bibfield  {author} {\bibinfo {author} {\bibfnamefont {B.}~\bibnamefont
  {Behin-Aein}}, \bibinfo {author} {\bibfnamefont {J.-P.}\ \bibnamefont
  {Wang}}, \ and\ \bibinfo {author} {\bibfnamefont {R.}~\bibnamefont
  {Wiesendanger}},\ }\href {\doibase 10.1557/mrs.2014.166} {\bibfield
  {journal} {\bibinfo  {journal} {MRS Bull.}\ }\textbf {\bibinfo {volume}
  {39}},\ \bibinfo {pages} {696} (\bibinfo {year} {2014})}\BibitemShut
  {NoStop}%
\bibitem [{\citenamefont {Handy}\ and\ \citenamefont
  {Gregory}(1950)}]{Gregory50_5049}%
  \BibitemOpen
  \bibfield  {author} {\bibinfo {author} {\bibfnamefont {L.~L.}\ \bibnamefont
  {Handy}}\ and\ \bibinfo {author} {\bibfnamefont {N.~W.}\ \bibnamefont
  {Gregory}},\ }\href@noop {} {\bibfield  {journal} {\bibinfo  {journal} {J.
  Am. Chem. Soc.}\ ,\ \bibinfo {pages} {5049}} (\bibinfo {year}
  {1950})}\BibitemShut {NoStop}%
\bibitem [{\citenamefont {Lee}\ \emph {et~al.}(2020)\citenamefont {Lee},
  \citenamefont {Kwon}, \citenamefont {Kim}, \citenamefont {Yoon},
  \citenamefont {Song}, \citenamefont {Lee}, \citenamefont {Choi},
  \citenamefont {Son},\ and\ \citenamefont {Won}}]{Won20_166447}%
  \BibitemOpen
  \bibfield  {author} {\bibinfo {author} {\bibfnamefont {C.}~\bibnamefont
  {Lee}}, \bibinfo {author} {\bibfnamefont {H.~Y.}\ \bibnamefont {Kwon}},
  \bibinfo {author} {\bibfnamefont {N.~J.}\ \bibnamefont {Kim}}, \bibinfo
  {author} {\bibfnamefont {H.~G.}\ \bibnamefont {Yoon}}, \bibinfo {author}
  {\bibfnamefont {C.}~\bibnamefont {Song}}, \bibinfo {author} {\bibfnamefont
  {D.~B.}\ \bibnamefont {Lee}}, \bibinfo {author} {\bibfnamefont {J.~W.}\
  \bibnamefont {Choi}}, \bibinfo {author} {\bibfnamefont {Y.-W.}\ \bibnamefont
  {Son}}, \ and\ \bibinfo {author} {\bibfnamefont {C.}~\bibnamefont {Won}},\
  }\href {\doibase 10.1016/j.jmmm.2020.166447} {\bibfield  {journal} {\bibinfo
  {journal} {J. Magn. Magn. Mater.}\ }\textbf {\bibinfo {volume} {501}},\
  \bibinfo {pages} {166447} (\bibinfo {year} {2020})}\BibitemShut {NoStop}%
\bibitem [{\citenamefont {Mermin}\ and\ \citenamefont
  {Wagner}(1966)}]{Wagner66_1133}%
  \BibitemOpen
  \bibfield  {author} {\bibinfo {author} {\bibfnamefont {N.~D.}\ \bibnamefont
  {Mermin}}\ and\ \bibinfo {author} {\bibfnamefont {H.}~\bibnamefont
  {Wagner}},\ }\href {\doibase 10.1103/PhysRevLett.17.1133} {\bibfield
  {journal} {\bibinfo  {journal} {Phys. Rev. Lett.}\ }\textbf {\bibinfo
  {volume} {17}},\ \bibinfo {pages} {1133} (\bibinfo {year}
  {1966})}\BibitemShut {NoStop}%
\bibitem [{\citenamefont {Xu}\ \emph {et~al.}(2018)\citenamefont {Xu},
  \citenamefont {Feng}, \citenamefont {Xiang},\ and\ \citenamefont
  {Bellaiche}}]{Bellaiche18_57}%
  \BibitemOpen
  \bibfield  {author} {\bibinfo {author} {\bibfnamefont {C.}~\bibnamefont
  {Xu}}, \bibinfo {author} {\bibfnamefont {J.}~\bibnamefont {Feng}}, \bibinfo
  {author} {\bibfnamefont {H.}~\bibnamefont {Xiang}}, \ and\ \bibinfo {author}
  {\bibfnamefont {L.}~\bibnamefont {Bellaiche}},\ }\href {\doibase
  10.1038/s41524-018-0115-6} {\bibfield  {journal} {\bibinfo  {journal} {Comp.
  Mater.}\ ,\ \bibinfo {pages} {57}} (\bibinfo {year} {2018})}\BibitemShut
  {NoStop}%
\bibitem [{\citenamefont {McGuire}(2017)}]{Mcguire17_121}%
  \BibitemOpen
  \bibfield  {author} {\bibinfo {author} {\bibfnamefont {M.~A.}\ \bibnamefont
  {McGuire}},\ }\href@noop {} {\bibfield  {journal} {\bibinfo  {journal}
  {Crystals}\ }\textbf {\bibinfo {volume} {7}},\ \bibinfo {pages} {121}
  (\bibinfo {year} {2017})}\BibitemShut {NoStop}%
\bibitem [{\citenamefont {Anderson}(1950)}]{Anderson50_350}%
  \BibitemOpen
  \bibfield  {author} {\bibinfo {author} {\bibfnamefont {P.~W.}\ \bibnamefont
  {Anderson}},\ }\href {\doibase 10.1103/PhysRev.79.350} {\bibfield  {journal}
  {\bibinfo  {journal} {Phys. Rev.}\ }\textbf {\bibinfo {volume} {79}},\
  \bibinfo {pages} {350} (\bibinfo {year} {1950})}\BibitemShut {NoStop}%
\bibitem [{\citenamefont {Kresse}\ and\ \citenamefont
  {Furthm\"uller}(1996{\natexlab{a}})}]{Furthmuller96_15}%
  \BibitemOpen
  \bibfield  {author} {\bibinfo {author} {\bibfnamefont {G.}~\bibnamefont
  {Kresse}}\ and\ \bibinfo {author} {\bibfnamefont {J.}~\bibnamefont
  {Furthm\"uller}},\ }\href {\doibase 10.1016/0927-0256(96)00008-0} {\bibfield
  {journal} {\bibinfo  {journal} {Comput. Mater. Sci.}\ }\textbf {\bibinfo
  {volume} {6}},\ \bibinfo {pages} {15} (\bibinfo {year}
  {1996}{\natexlab{a}})}\BibitemShut {NoStop}%
\bibitem [{\citenamefont {Kresse}\ and\ \citenamefont
  {Furthm\"uller}(1996{\natexlab{b}})}]{Furthmuller96_11169}%
  \BibitemOpen
  \bibfield  {author} {\bibinfo {author} {\bibfnamefont {G.}~\bibnamefont
  {Kresse}}\ and\ \bibinfo {author} {\bibfnamefont {J.}~\bibnamefont
  {Furthm\"uller}},\ }\href {\doibase 10.1103/PhysRevB.54.11169} {\bibfield
  {journal} {\bibinfo  {journal} {Phys. Rev. B}\ }\textbf {\bibinfo {volume}
  {54}},\ \bibinfo {pages} {11169} (\bibinfo {year}
  {1996}{\natexlab{b}})}\BibitemShut {NoStop}%
\bibitem [{\citenamefont {Bl\"ochl}(1994)}]{Blochl94_17953}%
  \BibitemOpen
  \bibfield  {author} {\bibinfo {author} {\bibfnamefont {P.~E.}\ \bibnamefont
  {Bl\"ochl}},\ }\href {\doibase 10.1103/PhysRevB.50.17953} {\bibfield
  {journal} {\bibinfo  {journal} {Phys. Rev. B}\ }\textbf {\bibinfo {volume}
  {50}},\ \bibinfo {pages} {17953} (\bibinfo {year} {1994})}\BibitemShut
  {NoStop}%
\bibitem [{\citenamefont {Perdew}\ \emph {et~al.}(1996)\citenamefont {Perdew},
  \citenamefont {Burke},\ and\ \citenamefont {Ernzerhof}}]{Ernzerhof96_3865}%
  \BibitemOpen
  \bibfield  {author} {\bibinfo {author} {\bibfnamefont {J.~P.}\ \bibnamefont
  {Perdew}}, \bibinfo {author} {\bibfnamefont {K.}~\bibnamefont {Burke}}, \
  and\ \bibinfo {author} {\bibfnamefont {M.}~\bibnamefont {Ernzerhof}},\ }\href
  {\doibase 10.1103/PhysRevLett.77.3865} {\bibfield  {journal} {\bibinfo
  {journal} {Phys. Rev. Lett.}\ }\textbf {\bibinfo {volume} {77}},\ \bibinfo
  {pages} {3865} (\bibinfo {year} {1996})}\BibitemShut {NoStop}%
\bibitem [{\citenamefont {Grimme}\ \emph {et~al.}(2010)\citenamefont {Grimme},
  \citenamefont {Antony}, \citenamefont {Ehrlich},\ and\ \citenamefont
  {Krieg}}]{Krieg10_132}%
  \BibitemOpen
  \bibfield  {author} {\bibinfo {author} {\bibfnamefont {S.}~\bibnamefont
  {Grimme}}, \bibinfo {author} {\bibfnamefont {J.}~\bibnamefont {Antony}},
  \bibinfo {author} {\bibfnamefont {S.}~\bibnamefont {Ehrlich}}, \ and\
  \bibinfo {author} {\bibfnamefont {S.}~\bibnamefont {Krieg}},\ }\href
  {\doibase 10.1063/1.3382344} {\bibfield  {journal} {\bibinfo  {journal} {J.
  Chem. Phys.}\ }\textbf {\bibinfo {volume} {132}},\ \bibinfo {pages} {154104}
  (\bibinfo {year} {2010})}\BibitemShut {NoStop}%
\bibitem [{\citenamefont {Keffer}(1961)}]{Keffer61_896}%
  \BibitemOpen
  \bibfield  {author} {\bibinfo {author} {\bibfnamefont {F.}~\bibnamefont
  {Keffer}},\ }\href {\doibase 10.1103/PhysRev.126.896} {\bibfield  {journal}
  {\bibinfo  {journal} {Phys. Rev.}\ }\textbf {\bibinfo {volume} {126}},\
  \bibinfo {pages} {896} (\bibinfo {year} {1961})}\BibitemShut {NoStop}%
\bibitem [{\citenamefont {Fert}\ \emph {et~al.}(2013)\citenamefont {Fert},
  \citenamefont {Cros},\ and\ \citenamefont {Sampaio}}]{Sampaio13_152}%
  \BibitemOpen
  \bibfield  {author} {\bibinfo {author} {\bibfnamefont {A.}~\bibnamefont
  {Fert}}, \bibinfo {author} {\bibfnamefont {V.}~\bibnamefont {Cros}}, \ and\
  \bibinfo {author} {\bibfnamefont {J.}~\bibnamefont {Sampaio}},\ }\href
  {\doibase 10.1038/nnano.2013.29} {\bibfield  {journal} {\bibinfo  {journal}
  {Nat. Nano.}\ }\textbf {\bibinfo {volume} {8}},\ \bibinfo {pages} {152}
  (\bibinfo {year} {2013})}\BibitemShut {NoStop}%
\end{thebibliography}%


\end{document}